\documentclass[runningheads]{svjour2}
\smartqed  
\usepackage{graphicx}
\newcommand{\zs}{Z$_{\odot}$}
\journalname{Astronomy and Astrophysics Review}
\begin{document}

\title{On the ``Galactic Habitable Zone''
}
\subtitle{}


\author{Nikos Prantzos     
}


\institute{
              Institut d'Astrophysique de Paris, UMR7095 CNRS\\
              Univ. P. \& M. Curie, 98bis, Bd Arago, 75014 Paris \\
              Tel.: 33 1 44 32 81 88, Fax: 33 1 44 32 80 01\\
              \email{prantzos@iap.fr} 
}

\date{{\it Strategies  for   Life  Detection},  ISSI  Bern,  April  24-28 2006  }

\maketitle

\begin{abstract}
The concept of Galactic Habitable Zone (GHZ) was  introduced a few years ago as an extension of the much older concept of Circumstellar Habitable Zone. However, the physical processes underlying the former concept are hard to identify and even harder to quantify. That difficulty does not allow us, at present, to draw any significant conclusions about the extent of the GHZ: it  may well be that the entire Milky Way disk is suitable for complex life.
\end{abstract}

\section{Introduction}
\label{intro}

The modern study of the ``habitability'' of  circumstellar environments started almost half a century  ago (Huang 1959). The concept of a Circumstellar Habitable Zone (CHZ) is relatively well defined, being tightly related to the requirement of the presence of liquid water as a necessary condition for life-as-we-know-it; the corresponding temperature range  is a function of the luminosity of the star and of the distance of the planet from it. An important amount of recent work, drawing on various disciplines (planetary dynamics, atmospheric physics, geology, biology etc.) refined considerably our understanding of various factors that may affect the CHZ; despite that progress, the subject should still be considered to be in  its infancy   (see.g. Chyba and Hand 2005 or Gaidos and Selsis 2005, and references therein).

Habitability on a larger scale was considered a few years ago, by Gonzalez et al. (2001) who introduced the concept of Galactic Habitable Zone (GHZ). The underlying idea is that various physical processes, which may favour the development or the destruction of complex life, may depend strongly on the temporal and spatial position  in the Milky Way. For instance, the risk of a  supernova explosion sufficiently close to represent a threat for life is, in general, larger in the inner Galaxy than in the outer one, and has been larger in the past than at present. Another example is offered by the metallicity (the amount of elements heavier than hydrogen and helium) of  the interstellar medium, which varies  across the Milky Way disk, and which may be  important for the existence of Earth-like planets. Indeed, the host stars of the $\sim$180 extrasolar giant planets detected so far are, on average, more metal-rich than stars with no planets in our cosmic neighborhood (e.g. Fischer and Valenti 2005). Several other factors, potentially important for the GHZ, are discussed in Gonzalez (2005).

The concept of GHZ is much less well defined than the one of CHZ, since none of the presumably relevant factors can be quantified in a satisfactory way. Indeed, the role of the metallicity in the formation and survival of Earth-like 
planets is not really understood at present, while the ``lethality'' of supernovae and other cosmic explosions is hard to assess. The first study attempting to quantitatively account for such effects is made by Lineweaver et al. (2004, herefater L04), with a detailed model for the chemical evolution of the Milky Way disk. They find  that the probability of having an environment favourable to complex life is larger in a ``ring'' (a few kpc wide) surrounding the Milky Way center and  spreading outwards in the course of the Galaxy's evolution. 

We repeat that exercice here with a model that reproduces satisfactorily the major observables of the Milky Way disk (Sec. 2). In Sec. 3 we discuss the role of metallicity and we quantify it in a different (and, presumably, more realistic) way than L04, in the light of recent simulations of planetary formation. In Sec. 4 we discuss the risk of SN explosions and conclude that it can hardly be quantified at present, in view of our ignorance of how robuste life really is. For comparison purposes, though,  we adopt the same risk factor as the one defined in L04. In Sec. 5 we present our results, showing that the GHZ may, in fact, extend to the whole Galactic disk today. We conclude that, at the present stage of our knowledge, the GHZ may extend to the entire MW disk.

\begin{figure*}
\centering
\includegraphics[angle=-90,width=\textwidth]{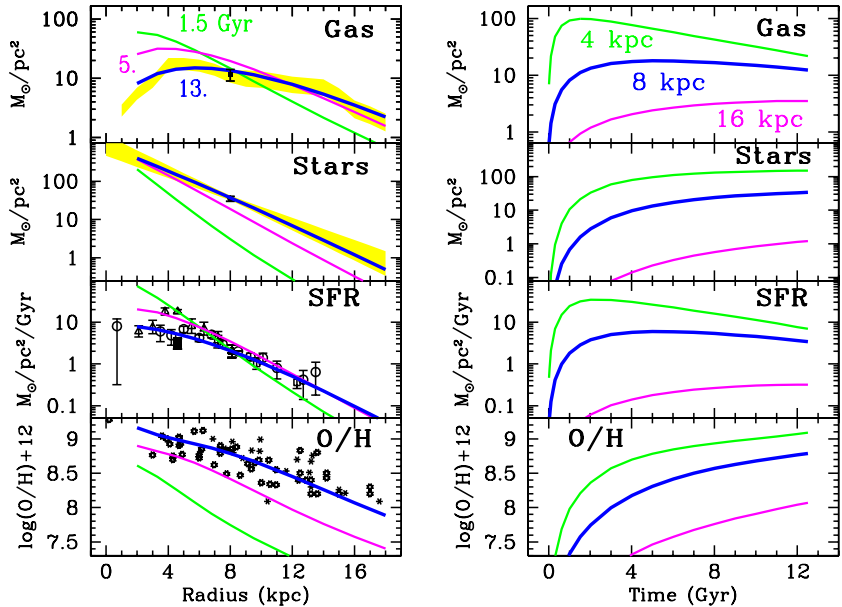}
\caption{The chemical evolution of the Milky Way disk, obtained in the framework of a one-dimensional model with radial symmetry (see Sec. 1 for a description of the model). {\it Left}, from top to bottom: radial profiles of the surface density of gas, stars, star formation rate (SFR), and of the oxygen abundance. Profiles are displayed at three epochs, namely at 1.5, 5, and 13 Gyr; the latter (thick curve in all figures) is compared to observations concerning the present-day MW disk (grey shaded areas in the first two figures, data points in the last two). {\it Right}, from top to bottom: the same quantities are plotted as a function of time for three different disk regions, located at galactocentric distances of 4, 8 and 16 kpc; the second one (thick curves in all panels of the right column) corresponds to the solar neighborhood. 
}
\label{fig:1}       
\end{figure*}

\section{The evolution of the Milky Way disk}

The evolution of the solar neighborhood is rather well constrained, due to the large body of observational data available. In particular, the metallicity distribution of long-lived stars suggests a slow formation of the local disk, through a prolonged period of gaseous infall (e.g. Goswami and Prantzos 2000; see however, Haywood 2006 for a different view). For the rest of the disk, available data concern mainly its current status and not its past history;  in those conditions, it is impossible to derive a unique evolutionary path. Still, the radial properties of the Milky Way disk (profiles of stars, gas, star formation rate, metallicity, colours etc.), constrain significantly its evolution and point to a scenario of "inside-out" formation (e.g. Prantzos and Aubert 1995).  
Such scenarii fit ``naturally'' in the currently favoured paradigm of galaxy formation in a cold dark matter Universe (e.g. Naab and Ostriker 2006).

Some results of a model developed along those lines (Boissier and Prantzos 1999, Hou, Prantzos and Boissier 2000) are presented in Fig. 1. Note that the evolution of the Galactic bulge (innermost $\sim$2 kpc) is not studied here, since it is much less well constrained than the rest of the disk; the conclusions, however, depend very little on that point.
The model assumes a star formation rate proportional to $R^{-1}$ (where $R$ is the galactocentric distance), inspired from the theory of star formation induced by spiral waves in disks. It satisfies all the available constraints for the solar neighborhood and the rest of the disk; some of those constraints appear in Fig. 1. Among the successes of the model, one should mention the prediction that the abundance gradient should flatten with time, due to the "inside-out" formation scheme (bottom right panel in Fig. 1), a prediction which is quantitatively supported by recent measurements of the metallicity profile in objects of various ages (Maciel et al. 2006).

It should be stressed, however, that the output of the model is only as good as the adopted observational constraints. For instance, there is still some uncertainty concerning the level of the oxygen abundance gradient: the "standard" value of dlog(O/H)/dR = -- 0.07 dex/kpc is challenged by the recent evaluations of Daflon and Cunha (2004), who find only half that value. If the latter turns out to be true, some of the model parameters (e.g. the adopted radial dependence of the infall rate and/or the star formation rate) should have to be revised.

In any case, it appears rather well established now that the star formation
rate and the metallicity are more important in the inner disk than in the outer one, and that this trend has been even stronger in the past (see left panel in Fig. 1). Some authors have suggested that those two parameters play an important role in the "habitability" of a given region of the Galactic disk (Gonzalez et al. 2001, L04); we discuss briefly that role in the next two sections.
 
\section{On the Probability of having stars with Earth-like planets}
\label{sec:2}

Since the first detection of an extrasolar planet around Peg 51 (Mayor and Queloz 1995), more than 170 stars in the solar neighborhood have been found to host planets (e.g. Butler et al. 2006). The masses of those planets range from 0.03 to 18 Jupiter masses and their distances to the host star lie in the range of 0.03-6 AU.  These ranges of mass and distance, however, are at present due to selection effects, resulting from  the current limitations of detection techniques. Continuous improvement of those techniques may well reveal the presence of smaller, Earth-like, planets at small distances from the host stars (as well as massive ones further away). In fact, the mass distribution of the detected planets is $dN/dM \propto M^{-1.1}$ (Butler et al. 2006) and suggests that Earth-like planets should be quite common (unless a yet unknown physical effect truncates the distribution from the low mass end). 

The unexpected existence  of    "Hot Jupiters"   is usually interpreted in terms of planetary migration (see e.g. Papaloizou and Terquem 2006 for a review): those gaseous giants can, in principle, be formed at a distance of several AU from their star (where gas is available for accretion onto an already rapidly formed rocky core); their subsequent interaction with the proto-planetary disk leads, in general,  to loss of angular momentum and migration of the planets inwards. On their way, those planets destroy the disk and any smaller planets that may have been formed there. Thus, the presence of Hot Jupiters around some stars implies that the probability of life (at least as we know it) in the corresponding stellar system is rather small.

\begin{figure}
\centering
\includegraphics[width=0.8\textwidth]{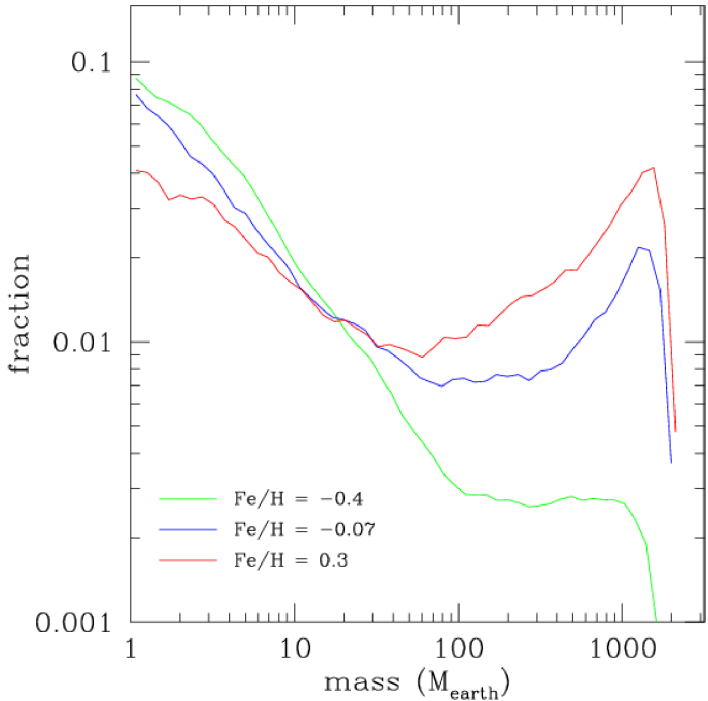}
\caption{Fraction of planets around a 1 M$_{\odot}$ star as a function of their mass (from simulations of Mordasini et al. 2006). Results are displayed for three different values of the metallicity of the star (and of the corresponding proto-planetary disk), as shown on the figure. It is seen that, high metallicities favour larger fractions of massive planets (in rough agreement with observations), while at low metallicities the presence of Earth-like planets is enhanced rather than suppressed. (Courtesy Y. Alibert). }
\label{fig:2}       
\end{figure}

\begin{figure}
\centering
\includegraphics[angle=-90,width=\textwidth]{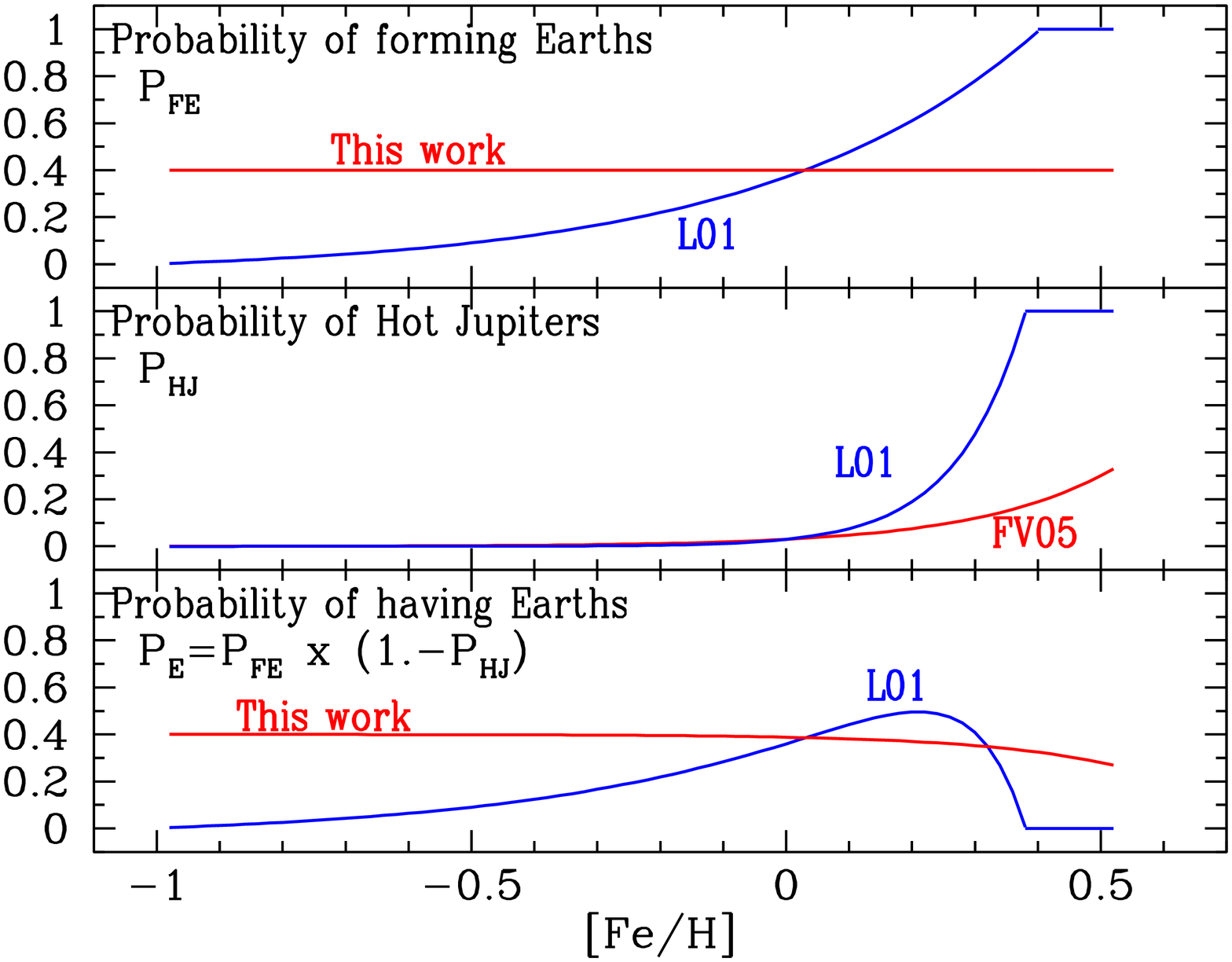}
\caption{Role of metallicity of the proto-stellar nebula in the formation and presence of Earth-like planets around solar-mass stars, according to various estimates. {\it Top:} The probability of forming Earths has been taken as proportional to metallicity in Lineweaver (2001, L01), while it is assumed to be quasi-independent of metallicity (at least for [Fe/H]$>$-1) here, after the results displayed in Fig. 2. {\it Middle:} The probability of forming Hot Jupiters (=destroying Earth-like planets in circumstellar habitable zones) is larger in absolute value and steeper as a function of metallicity in L01 than in Fischer and Valenti (2005, FV05); the latter function is adopted here. {\it Bottom:} The probability of having Earths, obtained from the 2 previous ones, in L01 and in this work. }
\label{fig:3}       
\end{figure}

A key feature of the stars hosting planets is their high metallicity (compared to stars wih no planets). That feature was first noticed by  Gonzalez (1997), who  intrepreted it in terms of accretion of (metal enriched) planetesimals onto the star. However, subsequent studies found no correlation of the stellar metallicity with the depth of the convective zone of the star, invalidating that idea (see e.g. Fischer and Valenti 2005, hereafter FV05, and references therein). Thus, it appears that the high metallicity is intrinsic to the star, and it presumably plays an important role to the giant planet formation. FV05 quantified the effect, finding that the fraction of FGK stars with Hot Jupiters  increases sharply with metallicity $Z$, as $P_{HJ}$ = 0.03 $ (Z/Z_{\odot})^2$, where $Z$ designates Fe abundance. Note that this function has lower values and is much less steep than the one suggested by Lineweaver (2001, hereafter L01), which reaches a value of $P_{HJ}$ = 0.5 at a metallicity of 2 \zs; the latter function is  used in the calculations of L04.

The impact of metallicity on the formation of Earth-like planets is unknown at present. Metals (in the form of dust) are obviously necessary for the formation of planetesimals, but the required amount depends on the assumed scenario (and its initial conditions, e.g. size of protoplanetary disk). Inspired by the early results on extrasolar planet host stars, several authors argued for an important role of metallicity. Thus, in L01 it is suggested that the probability of forming Earth-like planets should simply have a linear dependence on metallicity  $P_{FE} \propto Z$.  On the other hand, Zinnecker (2003) argues for a threshold on metallicity, of the order of 1/2 \zs, for the formation of Earth-sized planets.

However, the situation may be more complicated than the one emerging from simple analytical arguments. Recent (and yet unpublished) numerical simulations by the Bern group find that at low metallicities, the decreasing probability of forming giant planets leaves quite a lot of metals to form a significant number of Earth-sized planets (see Fig. 2). Those simulations cover a factor of $\sim$3 in metallicity (from $\sim$0.6 \zs \ to 2 \zs) and they roughly reproduce the observations concerning $P_{HJ} (Z)$ (see above); their predictions for $P_{FE}(Z)$ have to be confirmed by further simulations and, ultimately, by observations. They suggest, however, that the formation of Earth-like planets may be quite common, even at low metallicity environments like the outer Galaxy and the early inner Galaxy.

Assuming that $P_{FE}(Z)$ and $P_{HJ}(Z)$ are known, the probability of having stars with Earth-like planets (but not Hot Jupiters, which destroy them) is simply: $P_E(Z) \ = \ P_{FE}(Z) \ (1-P_{HJ}(Z))$. However, it is clear from the previous paragraphs that neither of the terms of the right member of that equation can be accurately evaluated at present\footnote{In fact, it is not even certain that the migration of Hot Jupiters inwards prohibits the existence  of terrestrial planets in the habitable zone: based on recent simulations of planetary system formation, Raymond et al. (2006) find that ``.. about 34\% of giant planetary systems in our sample permit an Earth-like planet of at least 0.3 M$_{Earth}$ to form in the habitable zone''.}. Just
for illustration purposes, we adopt in this work a set of metallicity-dependent probabilities different from the one adopted in L04 (who assumed $P_{FE}$ linearly dependent on $Z$, and $P_{HJ}$ strongly increasing with $Z$):  we assume that $P_{FE}=const$=0.4 (so that the metallicity integrated probability is the same as in L04) for $Z>$0.1 $Z_{\odot}$ and $P_{HJ}$ from FV05. The two sets appear in Fig. 3, which also displays $P_E$ (bottom panel): with our set, $P_E$   extends to non-zero values at metallicities much lower and higher than  in in the work of L04. From that factor alone it is anticipated that any GHZ (found through a chemical evolution model) will be much larger with our set of data than with the one of L04.

\section{On the probability of life surviving supernova explosions} 

The potential threat for complex life on Earth that represent nearby SN explosions was first studied by Ruderman (1974). He pointed out that energetic radiation from such events, in the form of hard X-rays, gamma-rays or cosmic rays, may (partially or totally) destroy the Earth's atmospheric ozone, leaving land life exposed to lethal does of UV fluxes from the Sun. The paper went completely unnoticed, with just one citation for about 20 years. In the last few years, however, a large number of studies were devoted to that topic (for reasons that are not quite clear to the author of this paper). Two factors are, perhaps, at the origin of that interest: the availability of complex models of Earth's atmospheric structure and chemistry; and the discovery of extrasolar giant planets. The latter suggests that Earth-like planets may also be abundant in the Galaxy; however, complex life on them may be a rare phenomenon, because of various cosmic threats like SN explosions. This is the basic idea underlying the concepts of GHZ (Gonzalez et al. 2001) and of the Rare Earth (Brownlee and Ward 2002): complex life in the Universe may be rare, not for intrinsic reasons (i.e. improbability of development of life on  a planet), but for extrinsic ones, related to the hostile cosmic environment.

Despite the simplicity of the idea, however, the studies devoted to the topic revealed that it is very difficult  to quantify the SN threat for life (or  any other cosmic threat). Studies of that kind evaluate the energetic particle flux/irradiation impinging on Earth, which  could induce a significant number of gene mutations (based on our understanding of  such mutations in various organisms on Earth). Knowing the intrinsic emissivity of energetic particles from a typical SN (which depends essentially on the configuration of the progenitor star and the energetics of the explosion), one may then calculate a lower limit for the distance to the SN, for such fluxes/irradiations to occur. The distribution of the rates of various SN types in the Milky Way (presumably known to within  a factor of a few) is then used to evaluate the frequency/probability  of such events in our vicinity and elsewhere in the Galaxy.

Numerous studies in the last few years explored several aspects of the scenario, with the use of models of various degrees of complexity for the atmosphere of the Earth (or Mars) and the transfer of high energy radiation through it (see e.g. Ejzak et al. 2006 and references therein).  Thus, Smith et al. (2004) find that a substantial fraction ($\sim$1  \%) of the high energy radiation (X- and $\gamma$- rays)  impinging on a thick atmosphere (column density $\sim$100 g cm$^{-2}$, like the Earth's) may reach the ground and induce biologically important mutations. Ejzak et al. (2006) find that the ozone depletion depends mainly on the total irradiation (for durations in the 10$^{-1}$ - 10$^8$ s range) and only slightly  on the received flux; this result allows one to deal with both short ($\gamma$-ray bursts\footnote{Gamma-ray bursts are more powerful, but also much rarer events than supernovae. Their beamed energy makes them lethal from much larger distances than SN (several kpc in the former case, compared to a few pc in the latter ) and several studies are recently devoted to that topic (Ejzak et al. 2006 and references therein). However, they are associated with extragalactic regions of low metallicity (in the few cases with available observations), implying that their frequency in the Milky Way has been probably close to zero in the past several Gyr; the formation of GRB progenitors in low metallicity environments is also favoured on theoretical grounds (e.g. Hirschi et al. 2005).} and long (SN explosions) events. They also, find that the overall result depends significantly on the shape of the spectrum (with harder spectra being more harmful).

Even if one assumes that such  calculations are realistic (which is far from being demonstrated), it is hard to draw any quantitative conclusions about the probability of {\it definitive sterilisation of a habitable planet} (which is the important quantity for the calculation of a GHZ). 
Even if 100\% lethality is assumed for all land animals after a nearby SN explosion, marine life will certainly survive to a large extent (since UV is absorbed from a couple of meters of water). In the case of the Earth, it took just a few hundred million years for marine life to spread on the land and evolve to dinosaurs and, 
ultimately, to humans; this is less than 4\% of the lifetime of a G-type star.
Even if land life on a planet  is destroyed from a nearby SN explosion, it may well reappear again after a few 10$^8$ yrs or so. Life displays unexpected robustness and a cosmic catastroph might even accelerate evolution towards life forms that are presently unknown\footnote{An illustration is offered by the ``Cambrian explosion'', with a myriad of complex life forms appearing  less than $\sim$40 Myr after the last ``snowball Earth'' (which presumably occured in the  Neoproterozoic era, 750 Myr ago).}. Only an extremely high frequency of such catastrophic events (say, more than one every few 10$^7$ yr) could, perhaps, ensure permanent disappearence of complex life from the surface of a planet.

Gehrels et al. (2003) find that significant biological effects due to ozone depletion, i.e. doubling of the  UV flux on Earth's surface, may arise for SN explosions closer than $D\sim$8 pc. Taking into account the estimated SN frequency in the Milky Way (a couple per century), the frequency of the Sun crossing spiral arms ($\sim$10 Gyr$^{-1}$), and the vertical to the Galactic plane density profile of the supernova progenitor stars (scaleheight $h_{SN}\sim$30 pc, comparable to the current vertical displacement of the Sun   $h_{\odot}\sim$24 pc), Gehrels et al. (2003) find that a SN should explode closer than 8 pc to the Sun with a frequency $f\sim$ 1.5 Gyr$^{-1}$. Repeating their calculation with a realistic radial density profile for SN resulting from massive stars (see profile of SFR in Fig. 1), we find  a number closer to $f\sim$1 Gyr$^{-1}$.
Note, however, the  sensitivity of that number to the assumed critical distance ($f\propto D^3$): reducing that distance from 8 pc to 6 pc, would correspondingly reduce the frequency to 0.4 Gyr$^{-1}$. In any case, it appears that no such catastrophic event occured close to the Earth in the past Gyr or so.

\begin{figure}
\centering
\includegraphics[angle=-90,width=0.6\textwidth]{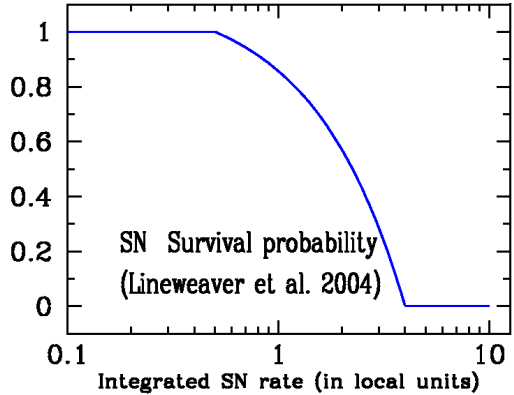}
\caption{The probability that complex life on an Earth-like planet survives a nearby supernova explosion, according to Lineweaver et al. (2004). It is plotted  as a function of the time-integrated supernova rate TIR$_{SN}$=$\int_t^{t+4 Gyr}$ $SNR(t') dt'$; the latter is expressed in units of the corresponding time-integrated SN rate in the solar neighborhood and in the last 4 Gyr. The probability is {\it quite arbitrarily} assumed to be $P=1$ for integrated SN rates less than 0.5 and $P=0$ for integrated SN rates larger than 4.}
\label{fig:4}       
\end{figure}

\begin{figure*}
\centering
\includegraphics[angle=-90,width=\textwidth]{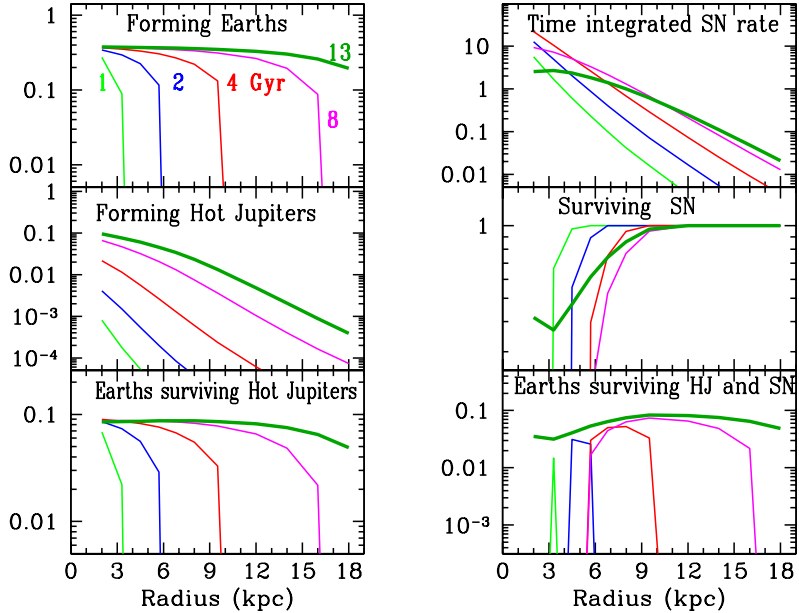}
\caption{Probabilities of various events as a function of galactocenric distance at five different epochs of the Milky Way evolution: 1, 2, 4, 8 and 13 Gyr, respectively (the latter is displayed with a {\it thick curve} in all figures). The probabilities of forming Earths $P_{FE}$ and Hot Jupiters $P_{HJ}$ and of actually having Earths $P_E=P_{FE} (1-P_{HJ})$ ({\it left part} of the figure) depend on the metallicity evolution (see Fig. 3), while the probability of life bearing planets surviving SN explosions is obtained with the criterion of Fig. 4. Finally, the overall probability for Earth-like planets with life ({\it bottom right} defines a ring in the MW disk, progressively migrating outwards; that "probability ring" is quite narrow at early times, but fairly extended today and peaking at about 10 kpc. }
\label{fig:5}       
\end{figure*}

In an attempt to circumvent the various unknowns, L04  quantify the risk for life represented by SN explosions by using the time integrated rate of SN within 4 Gyr (see Fig. 4). Adopting such a variable avoids considering the spiral arm passage and allows
for a rapid implementation of the SN risk factor in a a Galactic chemical evolution model; however,  the assignment of probabilities is rather arbitrary. For instance, it is assumed that $P<1$ for TIR$_{SN}$=1 local units, a rather strange assumption in view of the fact that life on Earth has well survived the local SN rate. Also, it is assume that $P=0$ for TIR$_{SN}$=4 local units, implying that if the local SN time integrated rate were just 4 times larger than 
what it has actually  been, then our planet would have been  permanently unable to host complex life. But such an increase in the SN rate would increase the frequency of ``lethal SN'' (closer than 8 pc, according to Gehrels et al. 2003), from 1 Gyr$^{-1}$ to 4 Gyr$^{-1}$, leaving still 250 Myr between lethal events, which is probably more than enough for a ``renaissance'' of land life. Assuming $P=0$ for substantially larger TIR values (say, for TIR$_{SN}$=10 local units) seems a safer bet, but it still constitutes a very imperfect evaluation of the SN threat.  For comparison purposes with L04, the ``SN risk factor'' of Fig. 4 is adopted also here.
  
\section{A GHZ as large as the whole Galaxy?}

Using the chemical evolution model of Sec. 2, the probabilities for forming Earth-like planets that survive the presence of Hot Jupiters of Sec. 3 and the risk factor from SN explosions from Sec. 4, we have calculated the distribution of stars potentially hosting Earth-like planets with complex life in the Milky Way.

The distribution of various probabilities as a function of galactocentric radius appears in Fig. 5, for five different epochs of the Galaxy's evolution, namely  1, 2, 4, 8 and 13 Gyr. Due to the inside-out formation scheme of the Milky Way, all metallicity-dependent probabilities peak early on in the inner disk and progressively increase outwards. The time-integrated SN rate is always higher in the inner than in the outer Galaxy. However, at late times its absolute value in the inner disk is smaller than at early times; as a result, the corrseponding probability for  surviving SN explosions, which is null in the inner disk at early times, becomes quite substantial at late times. 

The bottom line is depicted in the bottom right panel of Fig. 5. It shows that the overall probability of Earths surviving Hot Jupiters and SN has indeed a ring-like shape, quite narrow early-on (as it should, since there is basically no star formation in the outer disk at that time) and progressively  "migrating" outwards and becoming more and more extended. Today, that "ring"  peaks at $\sim$10 kpc, but it is quite large, since even the inner disk (the molecular ring) the SN risk factor is not much larger (just a factor of a few) than in the solar neighborhood. Obviously, the GHZ  extends to the quasi-totality of the galactic disk today. Fig. 6 (left panel) shows the same result as Fig. 5 (bottom right) in a different  way (space-time diagram), comparable with Fig. 3 of L04.

\begin{figure*}
\centering
\includegraphics[angle=-90,width=\textwidth]{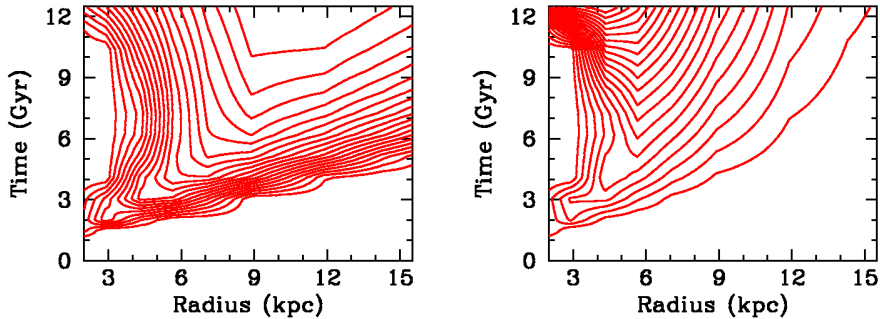}
\caption{  {\it Left:} Probability isocontours of having an Earth-like planet that survived SN explosions in the time-distance plane; as in the previous figure, that probability peaks today at around 10 kpc. {\it Right}: Multiplying that probability with the corresponding surface density of stars (much larger in the inner disk than in the outer one), one finds that it is more likely to find a star with an Earth-like planet that survived SN explosions in the inner Galaxy.}
\label{fig:6}       
\end{figure*}

In the right panel of Fig. 6, the radial probability of the left panel is multiplied with the corresponding number of stars created up to time $t$. Since there are more stars in the inner disk, this latter probability peaks in the inner Galaxy  (in other terms: the left panel displays the probability to have complex life around {\it one star} at a given position, while the right panel displays the probability of having complex life  {\it per unit volume} in a given position).  Thus, despite the high risk from SN early on in the inner disk, that place becomes later relatively "hospitable". Because of the large density of stars in the inner disk, it is more interesting to seek complex life there than in the outer disk; the solar neighbohood
(at 8 kpc from the center) is not privileged in that respect.

The results obtained in this section depend heavily on the assumptions made in Sec. 2 and 3, which are far from being well founded at present. It is clear that the contours of a GHZ in the Milky Way cannot be, even approximately, defined, either in space or in time; it may well be that most of our Galaxy is (and has been) suitable for life. Thus, the concept of a GHZ may have little or no significance at all. It should be considered, at best, as a broad framework, allowing us to  formulate our thoughts/educated guesses/knowledge about  a very complex phenomenon such as Life (origin, development and survival) in the Milky Way.

\begin{acknowledgements}
I am grateful to Franck Selsis for many useful discussions and a careful reading of the manuscript. 
\end{acknowledgements}



\end{document}